\documentstyle[PASJadd]{PASJ95}

\begin{document}

\title{Intracluster Supernova as a Possible Extra Energy Source
of Clusters}

\author{Shin {\sc Sasaki}\\
{\it Department of Physics, Tokyo Metropolitan University, Hachioji,
Tokyo 192-0397}\\
{\it E-mail: sasaki@phys.metro-u.ac.jp}}

\abst{The observed luminosity -- temperature relation of clusters is 
considerably steeper than that expected from a simple scaling relation.
Although extra energy input is a likely solution, its source has not
been identified.
We propose intracluster supernova as a possible source, and also
predict the supernova rate in intracluster space as a critical test.
}

\kword{ galaxies: clusters: general ---  (galaxies:) intergalactic
medium --- (stars:) supernovae: general --- }

\maketitle
\thispagestyle{headings}

\section{Introduction}

Clusters of galaxies are the largest virialized systems
in the universe, and their importance as a cosmological probe
is well-recognized.
If shock heating from gravitational collapse is the dominant mechanism
driving  intracluster medium evolution, the self-similar model
(e.g., Kaiser 1986) is a natural expectation. 
Based on the self-similar model, we can discuss the statistical
properties of clusters of galaxies, and most of the observational
results (for example, temperature function and mass -- temperature
relation) can be well explained.
However, some observational results are inconsistent
with the self-similar model (e.g., Evrard, Henry 1991; Kaiser 1991).
One of them is the luminosity -- temperature ($L$--$T$) relation.
In the self-similar model, the X-ray luminosity, $L_X $, can be
expressed as 
$ L_X \propto f_{\rm gas}^2 \left(1+z_{\rm vir}\right)^2 T^2 $,
where $z_{\rm vir}$ is the redshift of cluster virialization, 
$f_{\rm gas}$ is the gas mass fraction, and $T$ is the temperature.
This expression shows that if all clusters are formed at the same
redshift with the same gas mass fraction, the X-ray luminosity would be 
proportional to the temperature squared.
In fact, this relation is significantly steeper than that
observed (e.g., David et al. 1993).

In order to explain the $L$--$T$ relation, we need some physical
processes that can break the self-similar scaling.
Extra energy input is the most likely physical process
(e.g., Evrard, Henry 1991; Kaiser 1991).
Much literature has discussed cluster properties based on this
consideration, and shows that extra energy of
$ 0.1$ -- $1$ keV/particle is required to explain
the $L$ -- $T$ relation.

As a source of extra energy, supernovae (SNe) and active galactic
nuclei (AGN) are promising.
First, we briefly consider SNe.
The available energy from a supernova is $\epsilon \times 10^{51} {\rm 
erg}$, where $\epsilon$ is the conversion factor from the SN kinetic
energy to the thermal energy of the intracluster gas.
Assuming 1 supernova event per 100 $M_{\odot}$ of stars formed,
the total available energy is roughly expressed as
$ 10^{51} \epsilon ~ (M_{\rm gal}/100 M_{\odot}) ~ {\rm erg}$, 
where $M_{\rm gal}$ is the total mass of the stars in galaxies.
We then obtain the extra energy per particle as
$ 0.7 ~ \epsilon ~ [(M_{\rm gal}/M_{\rm gas})/0.2] ~ {\rm keV}$,
where $M_{\rm gas}$ is the mass of intracluster gas.
Here, we adopt $M_{\rm gal}/M_{\rm gas}=0.2$ as a typical value.
If $\epsilon$ is on the order of 1, the $L$--$T$ relation can be
explained by the energy input from SNe.
However, it is a serious problem that $\epsilon$ is on the order of 1,
since most of the SN kinetic energy converts to radiation energy.
For example, Thornton et al. (1998) have shown that about $ 90\%$ of the 
SN kinetic energy is lost by radiation in galaxies, that is,
$ \epsilon < 0.1$.
Thus, it is difficult for SNe to be the energy source.
Although AGNs (jets or winds) are another candidate, their energy
conversion efficiency is very uncertain.
Therefore, the energy source has not been identified.

In this paper, we reconsider SN as an energy source while exploring the
possibility that they occur in intracluster space.

\section{Intracluster Supernovae}

For considering SNe as an energy source, the radiative energy loss
is the core of the problem.
To resolve this, the SN kinetic energy must convert to thermal
energy before radiative cooling becomes efficient.
This is possible if a SN occurs in a low-density environment,
i.e. in intracluster space, since the radiative energy loss rate
is proportional to the density squared.
According to Thornton et al. (1998), radiative cooling begins to
be efficient at $1.4 \times 10^6 ~ {\rm yr}$ from a SN explosion,
with the surrounding gas density and temperature being $10^{-3} {\rm
cm^{-3}}$ and 1000 K, respectively.
Until this epoch, we can use the Sedov--Taylor model and find
that the shock velocity is about $100 ~ {\rm km~s^{-1}}$ at
$ t= 1.4 \times 10^6 ~ {\rm yr}$ (Thornton et al. 1998).  
This corresponds to the sound velocity of gas with $T \sim 10^5
{\rm K}$.
Thus, if the temperature of the surrounding gas is higher than $10^5
{\rm K}$, then all of the SN kinetic energy would convert to
thermal energy without any radiative loss.

That is to say, the problem is whether sufficient SN explosions occur in
intracluster space. 
First of all, we consider how many stars there are in intracluster space. 
Theoretically, as a consequence of the stripping of stars from
galaxies in clusters through the effect of fast encounters with
other galaxies (e.g. Moore et al. 1996) and the interaction of
the galaxies with the tidal field of the cluster, itself,
the existence of an intracluster star has been expected (e.g., Dressler
1984). 
This idea is supported by observations: for example,
Gregg and West (1998) reported on optical features explained as tidal
debris generated during galaxy-galaxy and galaxy-cluster
interactions in the Coma cluster. 
N-body simulations predict that the fraction of mass in the
intracluster stars to the total cluster stars is greater than 10\%
(e.g., Miller 1983).

Observationally, evidence of intracluster stars also exists.
Although Zwicky (1951) first claimed to detect excess light between the 
galaxies of the Coma cluster, subsequent searches have not been
conclusive. 
Recently we have obtained more direct evidence.
First, many intracluster planetary nebulae have been found in the Virgo 
cluster (Feldmeier et al. 1998) and the Fornax cluster
(Theuns, Warren 1997).
Second, Ferguson et al. (1998) have reported on intracluster
red-giant-branch stars in the Virgo cluster using the HST.
The estimated fraction of intracluster stars to the total number of
cluster stars is very uncertain.
It varies from $10 \%$ to $ 80 \%$, depending on the literature.
Here, we adopt a fraction of $ 50 \%$ as a typical value;
that is, the number of stars in intracluster space is roughly
equivalent to that in galaxies.

We can then estimate the extra energy by intracluster SNe
by assuming the same SN rate as that given above.
We obtain

\begin{equation}
 0.7 \epsilon \left(\frac{M_{\rm IC,*}/M_{\rm gal}}{1}\right)
\left(\frac{M_{\rm gal}/M_{\rm gas}}{0.2}\right) ~~~
{\rm keV/particle},
\end{equation}
where $M_{\rm IC,*}$ is the total mass of the intracluster stars.
Considering intracluster stars stripped from galaxies by
galaxy--galaxy and galaxy--cluster interactions, massive objects like
groups of galaxies may have already been formed when intracluster SNe
explode. 
It is then likely that the temperature of the surrounding gas is
higher than $10^5 {\rm K}$.
Thus, from the above discussion, the energy conversion efficiency,
$\epsilon$, is on the order of 1. 
Excess energy by intracluster SNe becomes $\sim$ 0.7 keV /
particle.
It is noted that this value is the same as that in the case of SNe in
galaxies, when assuming no radiative loss.

\section{Discussion}

In the previous section, we discussed intracluster SNe.
The most uncertain point is the assumption that the 
SN rate in intracluster space is the same as that in galaxies.
This depends on what morphological type galaxies mainly release their
stars to intracluster space by stripping.
In nearby clusters, most interacting galaxies are of the late type.
Thus, it seems reasonable to assume that stripped stars are relatively
young and that the SN rate in intracluster space is the same as that in 
galaxies.

In order to study the effect of intracluster SNe on the cluster
properties quantitatively, we need to perform numerical calculations.
As shown in the above estimation, the excess energy by intracluster SNe
is equivalent to that by SNe in galaxies, when assuming no radiative
loss, which may be incorrect.
Thus, under the assumption that the SN rate in intracluster space
is the same as that in galaxies,
it is expected that the intracluster SNe affect such cluster properties
as those in galaxies when assuming no radiative loss.
We can then use the previous results, which consider the cluster
properties with extra energy by SNe in galaxies without radiative
loss as a good approximation for this case.
For example, Wu et al. (1998) studied the effect of extra energy by SNe
on the $L$--$T$ relation and other properties using a semi-analytic
model. 
Their main result is that one can fit the $L$--$T$ relation and others by
including extra energies of $\sim 1$ keV/particle.
It is very possible that intracluster SNe can release such energy as
that discussed above.
Thus, intracluster SNe can be the source of extra energy in clusters.

In order to confirm that intracluster SNe are in fact the extra energy
source, we need to consider some points.
If intracluster SNe are the extra energy source, they would eject much
metal into intracluster space.
However, it has been considered that the metal in the intracluster gas
is supplied from elliptical/S0 galaxies based on the correlation between 
the iron mass, $M_{\rm iron}$, and the optical luminosities in
elliptical/S0 galaxies, $L_{\rm E+S0}$ (e.g., Arnaud et al. 1992).
Although the origin of galaxy morphology remains an unsettled problem,
galaxy interactions, such as mergers, may be important processes in
elliptical-galaxy formation.
If so, it is expected that there would be a positive correlation between
the total mass of elliptical galaxies and that of intracluster stars,
since both elliptical galaxies and intracluster stars are formed through
galaxy interactions.
Thus, qualitatively, the observed correlation between $M_{\rm iron}$ and 
$L_{\rm E+S0}$ does not contradict this paper's scenario.
On the other hand, some literature claim that the required energy to
explain the $L$--$T$ relation is larger than $\sim 1$ keV/particle,
$2$ -- $3.5$ keV/particle, depending on the epoch of energy input,
the source distribution and so on (e.g., Wu et al. 1999; Loewenstein
2000). 
If their result is correct, the energy input from intracluster SNe
would explain only part of the required energy, and other energy sources
could be needed.  
In order to study these points and obtain to quantitative results, we
now wish to undertake a detailed study using a semi-analytic model.

At the present, the fraction of the intracluster stars to the total
cluster stars, which is an important value in this paper's scenario, is
very uncertain.  
If the fraction is much smaller than $\sim 50 \%$, intracluster SNe 
cannot release the required energy to explain the $L$--$T$ relation, and
other energy sources are needed.
The main reason for this uncertainty is that the total area of the
observed fields is small. 
If future surveys obtain more precise values, we could judge clearly
whether intracluster SNe are an extra energy source or not.

Until now, no SN has been discovered in intracluster space.
If stripping occurs only at a high redshift, type-II SNe in
intracluster space would also occur only at a high redshift.
Thus, we do not observe them in intracluster space at the present.
However, it is expected that type-Ia SNe occur, even at present.
Using the type-Ia SN rate in elliptical galaxies
[$0.98 h^2/(10^{10} L_B 100 {\rm ~yr})$,
van den Bergh, Tammann 1991] and assuming a
mass-to-luminosity ratio of 5, typical for the stellar content of an
old population, the type-Ia SN rate in intracluster space is

\begin{equation}
 0.08 \left(\frac{h}{0.65}\right)^2
\left(\frac{M_{\rm IC,*}}{10^{13}M_{\odot}}\right)
\left[\frac{M/L}{5(M/L)_{\odot}}\right]^{-1} ~~~ {\rm yr^{-1}},
\end{equation}
where $H_0=100 h ~ {\rm km~s^{-1}Mpc^{-1}}$ is the Hubble constant.
In rich clusters, like the Coma cluster, a type-Ia SN occurs 
every ten years.
This is a crucial test of the scenario that intracluster SNe are the
extra energy source.

\par
\vspace{1pc}\par
We would like to thank an anonymous referee for his/her useful comments
and suggestions.
This research was supported by the Grant-in-Aid for Scientific Research
from the Ministry of Education, Science, Sports and Culture of Japan (No.
11740157 and No. 12304009).

\section*{References}
\small

\re
Arnaud M., Rothenflug R., Boulade O., Vigroux L., Vangioni-Flam E.\
1991, A\&A 254, 49 

\re
David L.P., Slyz A., Jones C., Forman W., Vrtilek S.D.\ 1993, ApJ 412,
479 

\re
Dressler A.\ 1984, ARA\&A 22, 185

\re
Evrard A.E., Henry J.P.\ 1991, ApJ 383, 95

\re
Feldmeier J.J., Ciardullo R., Jacoby G.H.\ 1998, ApJ 503, 109

\re
Ferguson H.C., Tanvir N.R., von Hippel T.\ 1998, Nature 391, 461

\re
Gregg M.D., West M.J.,\ 1998, Nature 396, 549

\re
Kaiser N.\ 1986, MNRAS 222, 323

\re
Kaiser N.\ 1991, ApJ 383, 104

\re
Loewenstein M.\ 2000, ApJ 532, 17

\re
Miller G.E.\ 1983, ApJ 268, 495 

\re
Moore B., Katz N., Lake G., Dressler A., Oemler A. Jr 1996, Nature 379, 613 

\re
Theuns T., Warren S.J.\ 1997, MNRAS 284, L11

\re
Thornton K., Gaudlitz M., Janka H.-Th., Steinmetz M.\ 1998, ApJ 500,
95 

\re
van den Bergh S., Tammann G.A.\ 1991, ARA\&A 29, 363

\re
Wu K.K.S., Fabian A.C., Nulsen P.E.J.\ 1998, MNRAS 301, L20

\re
Wu K.K.S., Fabian A.C., Nulsen P.E.J.\ 1999, astro-ph/9907112

\re
Zwicky F.\ 1951, PASP 63, 61

\label{last}
\end{document}